# Wildfire Smoke: A Rigorous Challenge for HVAC Air Filters

Tanya Shirman,[1] Hediyeh Zamani[1] and Sissi Liu[1]

[1]Metalmark Innovations, PBC, Cambridge, MA, USA 02138

**Abstract**

Wildfires pose a significant air quality challenge as the smoke they produce can travel long distances and infiltrate indoor spaces. HVAC filters serve as a primary defense against this threat. In our study, we tested over 17 different types of commercially available filter media, including charged and uncharged synthetic, as well as fiberglass media, from leading global manufacturers, to assess their efficiency in removing smoke, using pine needle smoke as a proxy for wildfire smoke. Our findings revealed that charged polymer media, across all tested MERV grades, showed a rapid decline in smoke removal efficiency within minutes while causing no increase in airflow resistance. In contrast, mechanical media demonstrated more stable performance but experienced an increase in airflow resistance. Using scanning electron microscopy (SEM), we examined smoke deposition on fiber media, revealing unique interactions that contribute to the decline in filtration performance. These results suggest the need to reconsider current standards–whose testing is centered around inorganic particle removal–and the applicability of their ratings to smoke filtration. Our study also highlights the critical need to develop standard testing methods and evaluation criteria specific to smoke particles in HVAC filters to better protect human occupants during wildfire events.

**Keywords**

Wildfire smoke, PM2.5, HVAC filter media, filtration efficiency, filter morphology

**Synopsis**

Wildfire smoke is driving a paradigm shift in air quality, necessitating a new standard for the efficiency and performance of filter media.

1. **Introduction**

Over recent decades, the Western US and Canada have experienced a marked increase in the frequency and intensity of wildfires, posing escalating environmental and health threats.[1,2] Studies have shown that wildfire smoke can affect the air quality, visibility, and atmospheric chemistry of places that are hundreds, or even thousands, of kilometers away from the original locations of wildfires.[3] Wildfires emit a wide range of pollutants, including fine particulate matter (PM) and volatile organic compounds (VOCs), with studies showing that most wildfire PM is sub-micron and composed primarily of organic compounds (over 90%), along with elemental carbon and other inorganic elements.[1,2,4-6] Exposure to wildfire smoke poses serious health risks, particularly for respiratory and cardiac conditions, with hospital visits spiking during smoke events.[7-9] Recent studies suggest that wildfire smoke PM2.5 is linked to a 21% increased risk of developing dementia, more so than other air pollutants.[10] It may also impact cognition, leading to lower test scores and reduced future income.[11]

Buildings are viewed as safe shelters from smoke, however, due to their size, smoke PM can infiltrate indoors via HVAC systems and building envelopes.[12] Ensuring safe indoor environments is crucial for climate resilience, making HVAC filters essential for removing PM in residential and commercial spaces.

Most common HVAC filters in use today have a Minimum Efficiency Reporting Value (MERV) of 8-10, per ASHRAE Standard 52.2, chosen based on building mechanical system design parameters, HVAC system configurations, and energy considerations. To address wildfire smoke, upgrading HVAC filters to MERV 13 is typically recommended. Most filters are either mechanical or charged/electret. Increasingly, the latter is becoming more widespread, driven by their low cost and energy-saving benefits associated with their low airflow resistance. Only a limited number of studies have examined or characterized the impact of wildfire smoke on filter performance, leaving this area largely understudied. For example, Holder et al. found that the CADR of air cleaners with charged filters dropped by 95% with only a 7% increase in pressure after just 10 mg of pine needle (PN) smoke deposition.[13]

In our previous study, we analyzed the effect of PN smoke on the filtration efficiency of MERV 11 charged and fiberglass media.[14] In this study, we expanded our assessment to include 17 types of commercially available filter media, covering both charged (electret) and non-charged (mechanical) polymeric, as well as fiberglass media, with MERV ratings from 10 to 14. We measured the filtration efficiency of media samples after various durations of smoke deposition and analyzed the changes in the morphology of the filter media over time using Scanning Electron Microscopy (SEM). The findings highlight the performance challenges of common HVAC filters in protecting against wildfire smoke and aid in developing more appropriate smoke testing methods for filter media and filters. Equally important, they suggest more studies are needed to ensure comprehensive public health recommendations for preserving indoor air quality and protecting human occupants during wildfire events.

2. **Materials and Methods**

*Filter media.* All filter media tested were sourced directly from four globally recognized filter media manufacturers. In this paper, we refer to them as Manufacturer A, Manufacturer B, and so forth, in no particular order and without revealing their specific origins. We tested 17 different filter media types including charged (electret) polymer, uncharged polymer, and fiberglass media. The detailed information for each of the filter media samples is provided in Supporting Information (**SI, Table S1**). The minimum efficiency reported values (MERV) were in the range of 10 to 15 based on the manufacturer's specifications. All media types were tested for filtration efficiency using new, clean media samples. At least three samples of each type were measured.

*Filtration Efficiency Test Setup.* The benchtop smoke filtration testing setup was custom built and its details are described in SI (**Figure S1**). The air composition was analyzed using PM counting instruments, including an optical particle sizer (OPS Model 3330, TSI) capable of counting particles with a diameter in the range of 365 nm to 10 μm in the concentration range of 0–3,000 particles/cm$^3$, and a Scanning Mobility Particle Sizer (Nanoscan SMPS, Model 3910, TSI) capable of counting particles in the 10–365 nm size range at concentration range of 100–1,000,000 particles/cm$^3$. The system operates in the range of 2-15 LPM or 5.5 - 40 cm/s across a ~5 cm$^2$ flow exposed sample area with a corresponding pressure drop range of 40-500 Pa, depending on the media type. For all experiments, the temperature and relative humidity (RH) were monitored (SEN55, Sensirion, Stäfa, Switzerland) inside the smoke chamber and stayed within the ranges of 24–28 °C and 45–65% RH, respectively. The setup's design incorporates programable automated switching between the four conduit arms (samples and bypass), enabling a rapid test throughput, precision, and accuracy that helped to establish robust testing procedures for PN smoke removal efficiency.

*Experimental procedure, data collection, and analysis.* In a typical experiment, samples were cut into ~5 cm diameter (~2 inch) circles and placed into sample holders. The cross-sectional area of the sample media exposed to the air stream was approximately 5 cm$^2$ (Figure S1, D, brown circle). For each experimental run, a predetermined weight of PN was cut into approximately 1 cm long pieces, then combusted using a portable smoke infuser directly into the smoke chamber (**Figure S1, C**). The smoke in the chamber was given approximately 15 minutes to equilibrate, during which the benchtop system operated in bypass mode to allow the signal to stabilize. This was followed by a filtration efficiency test, during which air was passed through the first sample (channel 1) for 2 minutes, followed by another 2 minutes in bypass mode. This cycle was repeated for another two samples (channels 2 and 3). During this experiment, the air was continuously analyzed with the particle counter instruments. All of the filter media were tested at a flow rate of 10 L/min, corresponding to a face velocity of ~33 cm/s or ~60 fpm. This airflow velocity falls within the range of the flow velocities of a pleated HVAC filter's cross-section in practical applications.[15,16] For smoke deposition experiments, the smoke was allowed to pass through the samples for a predetermined amount of time (e.g. 20 min, 80 min) and then the filtration efficiency test was repeated. A more detailed description of the test procedure is provided in SI (**Figure S2**). Following data collection, the filtration efficiency of the tested media samples was calculated using Equation (1), where $C_{filter}$ is the average of the signals recorded during fltration mode and $C_{bypass}$ is the average of the signals recorded in the bypass mode immediately before and after the filtration mode.

Eq. 1 $\quad FE\% = (1 - C_{filter}/C_{bypass}) \times 100\%$

*Filter media characterization (SEM).* The filter morphology was analyzed using Field Emission Scanning Electron Microscopy (FE SEM, Gemini 360, Zeiss). The samples for SEM images were prepared by cutting filter media into small pieces of ~0.5 x 0.5 cm and placing them on an SEM holder.

3. Results and Discussion

*Smoke generation and characterization.*

PN smoke is considered a suitable model analyte for wildfire smoke and has been used by scientists at the U.S. EPA as well as other researchers in laboratory studies as a representative substance for biomass burning.[17] It is a challenging analyte characterized by complex composition and highly dynamic particle size distribution that depend on combustion temperature, concentration, and aging. There is currently no standard protocol for testing the filtration efficiency of filter media with PN smoke. Our criteria for selecting the appropriate conditions included ensuring the stability of the particle size distribution and PM concentration within the instrument's sensitivity range over the experiment duration. No particles larger than 400 nm were detected using a combination of OPS and SMPS. Hence, only SMPS was used in filtration efficiency experiments (**Figure S3, A**). Varying the weight of PN (from 30 to 350 mg) had little impact on the PM size distribution after the equilibration period, consistently showing a mono-modal size distribution between 50–400 nm, with a median diameter ranging from 100 to 160 nm. For this study, we opted to use 125 mg of PNs, which produces around 400,000 counts/cm³ (~1500 μg/m³) of particles within the 60-380 nm range, with a median diameter of ~160 nm (**Figure S3, B-C**). This size range is consistent with real-world conditions. As wildfire plume undergoes atmospheric transformations (ages), particle size has been found in the range of 100-300 nm.[18-21] Following the equilibration phase, the smoke particle size distribution remained reasonably stable for the experiment for approximately 1h (**Figure S3, C**).

*Filtration Efficiency and Pressure Drop.*

Seventeen types of filter media with MERV ratings ranging from 10 to 15, sourced from four different manufacturers (designated as Manufacturer A, B, C, and D), were used to test the temporal filtration efficiency of PN smoke and changes to pressure drop (PD). Their filtration efficiency across different smoke deposition periods was analyzed by averaging the size-resolved efficiency over particle sizes between 116 and 365 nm, as the efficiency followed almost linear trend across this size range (**Figure S4-6** and **Table S2**). The summary of the effects of smoke deposition on filtration efficiency per the associated MERV rating, media type, and manufacturer is shown in **Figure 1** and summarized in **Table 1**. In general, initial filtration efficiencies (~ 5 min of smoke deposition) of PN smoke of charged media were in agreement with the efficiencies as represented by the manufacturer-provided MERV ratings, while mechanical media demonstrated lower PN removal efficiency than their MERV ratings suggest (**Table 1**). Since PN smoke particles are smaller than 1 micron, only the $PM_{1.0}$ size removal efficiency associated with the MERV rating requirements is shown in the table, as it is the most relevant PM size category for smoke filtration. Noticeable variations in initial filtration efficiency were observed among media with the same

MERV grade from different manufacturers, likely due to differences in fabrication processes and media structure. The most conspicuous difference was observed for Manufacturer B's MERV 13 filter media (13-nano) which exhibited a steep decline in efficiency after a short period of smoke deposition (**Figure 1**). This filter media features a composite structure with nano- and micrometer fibers. While nanofibrous media is recognized for boosting filtration efficiency without increasing pressure drop, contributing to its growing popularity in new filter designs, this advantage does not hold for smoke deposition. Notably, the media showed an immediate and precipitous drop in efficiency, likely due to complex interactions between the smoke and nanofibers, which will be explored further below. On the other hand, charged synthetic media demonstrated higher initial removal efficiency of PN smoke than mechanical media of the same grade. As previously noted, the initial efficiency of mechanical media is generally lower than anticipated based on their MERV rating. This is particularly intriguing because, in this study, the face velocity was almost 6 times higher than the velocities used in standard testing for determining MERV ratings (~33 cm/s vs. 5.32 cm/s). The general performance characteristics of mechanical media for inorganic particle removal suggest an increase in efficiency as face velocity increases. However, this was not the case for smoke particles. Quite the opposite pattern was observed: the higher the face velocity, the lower the smoke filtration efficiency. The initial PD of charged media was significantly lower than that of mechanical media. As expected, the PD increases with the MERV rating of media, while higher filtration efficiency is also observed in the media (**Table 1**).

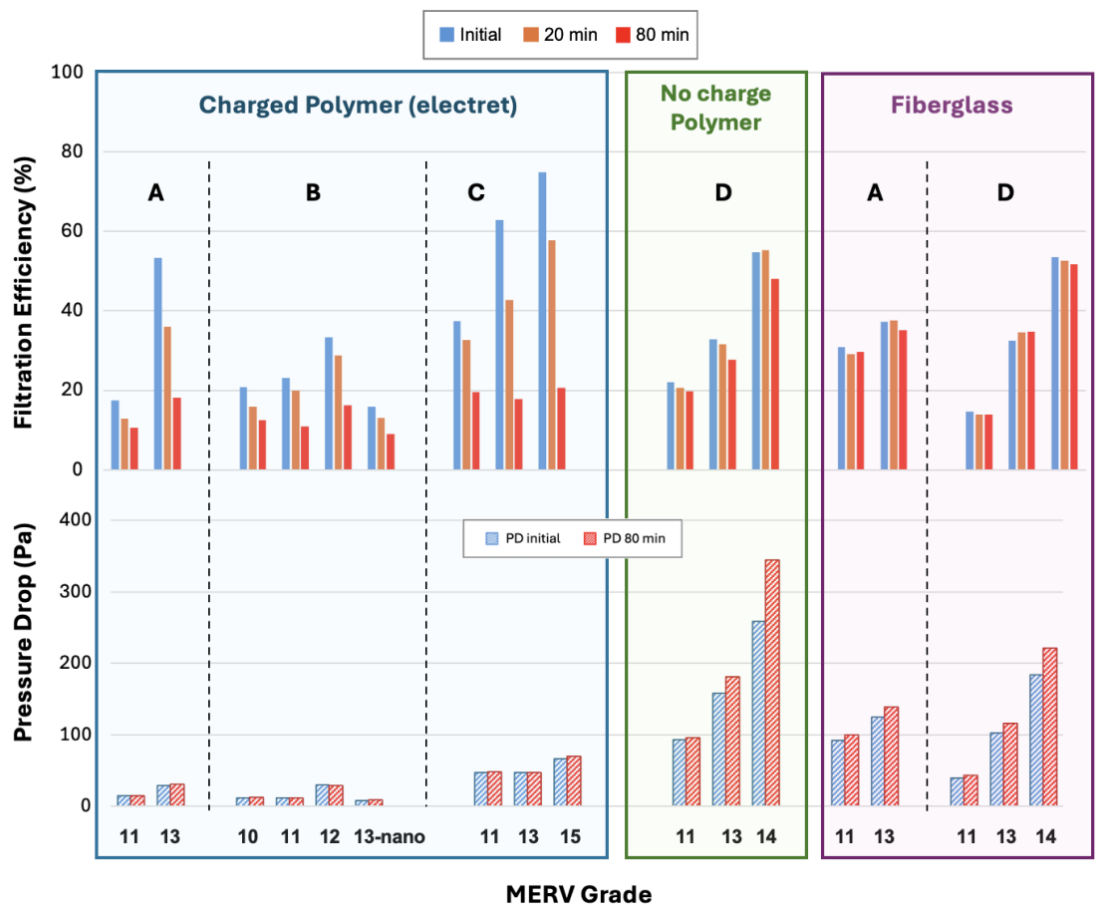

**Figure 1.** Summary of temporal filtration efficiency and pressure drop (PD) of various filter media measured during pine needle (PN) smoke deposition. Letters A-D designate different filter media manufacturers. Top panel: filtration efficiency of PN smoke measured after ~5 (initial, blue bars), 20 (orange bars), and 80 min (red bars) of continuous smoke deposition as measured using SMPS. Bottom panel: pressure drop measured across various samples after ~5 (initial, patterned blue bars) and 80 min (patterned red bars) of continuous smoke deposition. Experiments: flow rate 10L/min; at least three samples of each type were tested; the filtration efficiency is calculated based on an average over 5 particle size bins: 116, 154, 205, 274, 365 nm (See SI for more details).

Filter media samples were subjected to continuous PN smoke deposition for varying durations, and their changes in filtration efficiency and pressure drop were assessed after specific time intervals of smoke exposure, including after 20 and 80 minutes (Figure 1, orange and red bars, respectively, and Table 2). The PN smoke removal efficiency of charged synthetic media deteriorates rapidly with continuous deposition of PN smoke. For instance, the efficiency of MERV 13 and MERV 15 media from Manufacturer C, which initially showed the highest filtration efficiency, decreased by 72% after 80 minutes. In contrast, mechanical media, both uncharged polymer (synthetic) and fiberglass, demonstrated much higher stability toward PN smoke deposition in terms of filtration efficiency. For most mechanical media samples, the decrease in efficiency was less than 10% after 80 min of continuous deposition. The change in PD was insignificant for charged synthetic media regardless of the deposition time.

**Table 1**. Comparison of initial filtration efficiency, pressure drop of various samples, and their smoke deposition effect. Summary of the measured PN smoke filtration efficiency after various deposition periods, corresponding pressure drop (PD). Experimental conditions: 125 mg of PN, PM total counts ~ 400,000 #/cm$^3$ (~1500 µg/m$^3$), flow ~ 10 LPM, ~33 cm/s, filtration efficiency averaged over 50-300 nm PM size range.

| M[1] | MERV[2] | Pressure Drop, PD (Pa) | | Increase in PD (%) | Filtration Efficiency, FE[3] (%) | | Decrease in FE (%) | MERV Efficiency Range for PM$_{1.0}$[4] (%) |
|---|---|---|---|---|---|---|---|---|
| | | Initial | 80 min | | Initial | 80 min | | |
| **Charge Polymer (electret)** | | | | | | | | |
| A | 11 | 15 | 15 | 0 | 17.5 | 10.5 | 40 | 20 - 34 |
| | 13 | 29 | 31 | 7 | 53.3 | 18.1 | 66 | 50 - 74 |
| B | 10 | 12 | 13 | 8 | 20.8 | 12.5 | 40 | NA |
| | 11 | 12 | 12 | 0 | 23.1 | 10.9 | 53 | 20 - 34 |
| | 12 | 30 | 29 | -3 | 33.3 | 16.3 | 51 | 35 - 49 |
| | 13-nano | 8 | 9 | 13 | 15.8 | 9.0 | 43 | 50 - 74 |
| C | 11 | 47 | 48 | 2 | 37.5 | 19.7 | 47 | 20 - 34 |
| | 13 | 47 | 47 | 0 | 62.9 | 17.8 | 72 | 50 - 74 |
| | 15 | 66 | 70 | 6 | 74.8 | 20.6 | 72 | 85 - 94 |
| **Non-charged Polymer** | | | | | | | | |
| D | 11 | 93 | 96 | 2 | 22 | 19.7 | 10 | 20 - 34 |
| | 13 | 158 | 181 | 15 | 32.2 | 27.7 | 14 | 50 - 74 |
| | 14 | 259 | 345 | 33 | 54.7 | 48.1 | 12 | 75-85 |
| **Fiberglass** | | | | | | | | |
| A | 11 | 92 | 100 | 6 | 30.8 | 29.6 | 4 | 20 - 34 |
| | 13 | 125 | 139 | 11 | 37.2 | 35.2 | 5 | 50 - 74 |
| D | 11 | 40 | 43 | 8 | 14.6 | 13.9 | 5 | 20 - 34 |
| | 13 | 103 | 116 | 13 | 32.4 | 34.7 | -7 | 50 - 74 |
| | 14 | 184 | 221 | 20 | 53.5 | 51.8 | 3 | 75-85 |

[1]Manufacturer (M). [2]Minimum Efficiency Reporting Value (MERV). [3]Smoke particles removal efficiency measured at 33 cm/sec (60 fpm) of airflow. [4]ASHRAE 52.2 MERV Efficiency Range for PM$_{1.0}$ (0.3 – 1.0 µm).

It is generally believed that upgrading HVAC filters to higher MERV ratings improves protection against wildfire smoke. However, this does not apply to charged media. To summarize, regardless of their MERV ratings, 1) charged media's smoke removal efficiency dropped below 20% quickly; 2) post continuous smoke deposition, only a slight increase in PD was observed (PD remained nearly unchanged after 80 minutes, except for the 13-nano sample, which saw a 13% increase).

Mechanical media, on the other hand, showed a PD increase proportional to the MERV rating, with the highest being a 33% increase for the MERV 14 uncharged polymer (Manufacturer D).

*Fiber morphology after smoke deposition.*

Most studies have focused on the filtration efficiency of single-composition particles, such as solid inorganic particles or oil aerosols like dioctyl phthalate (DOP), resulting in a poor appreciation of the deposition behavior of mixed particles such as wildfire smoke particles. Understanding the interactions between smoke particulates and filter media fibers, including their deposition behaviors and resulting fiber morphology, is crucial for assessing the overall performance of air filters. Studies utilizing electron microscopy for characterizing challenged air filter surfaces are scarce.[14] To gain insights into the effect of PN smoke deposition on filter media morphology, the three representative types of media samples were characterized using SEM. **Figure 3** shows SEM images of charged synthetic and mechanical media (uncharged synthetic and fiberglass) from Manufacturers C and D, respectively, after various smoke deposition periods. All clean samples exhibit a smooth fiber surface morphology. However, after a brief period of initial filtration, deposits or film-like coatings appear on the fiber surfaces of all sample types, with charged media showing notably more coverage. After 80 minutes of deposition, all samples display droplet-like deposits resembling a "beads on a string" morphology on their surfaces. This phenomenon was also noted in our previous studies of PN smoke filtration.[14] The morphology of smoke particles on the fibers differs significantly from that of inorganic salt particles (e.g., KCl), which are characterized by dendritic structures. While droplet-like morphologies have also been reported in several studies focusing on liquid-phase organic aerosols, it is important to remember that filtration efficiency and pressure drop for mixed particles such as wildfire smoke depend on factors like particle size, charge, and composition.[22,23] The "beading" effect appears in all types of media that contain thin fibers with sub-micron diameters, and it forms even at the early stages of deposition. Moreover, his effect is observed across all media types, regardless of charge state or composition. It seems that the fiber aspect ratio plays a crucial role in interactions with smoke particles. In the case of composite media incorporating nanofibers (13-nano, Manufacturer B), the most extensive beading among all samples was observed even after a short deposition period (**Figures S7**). As mentioned previously, this media demonstrated an immediate drop in the removal efficiency to below 10% within the first few minutes of smoke filtration (**Figure 1, Table 2**).

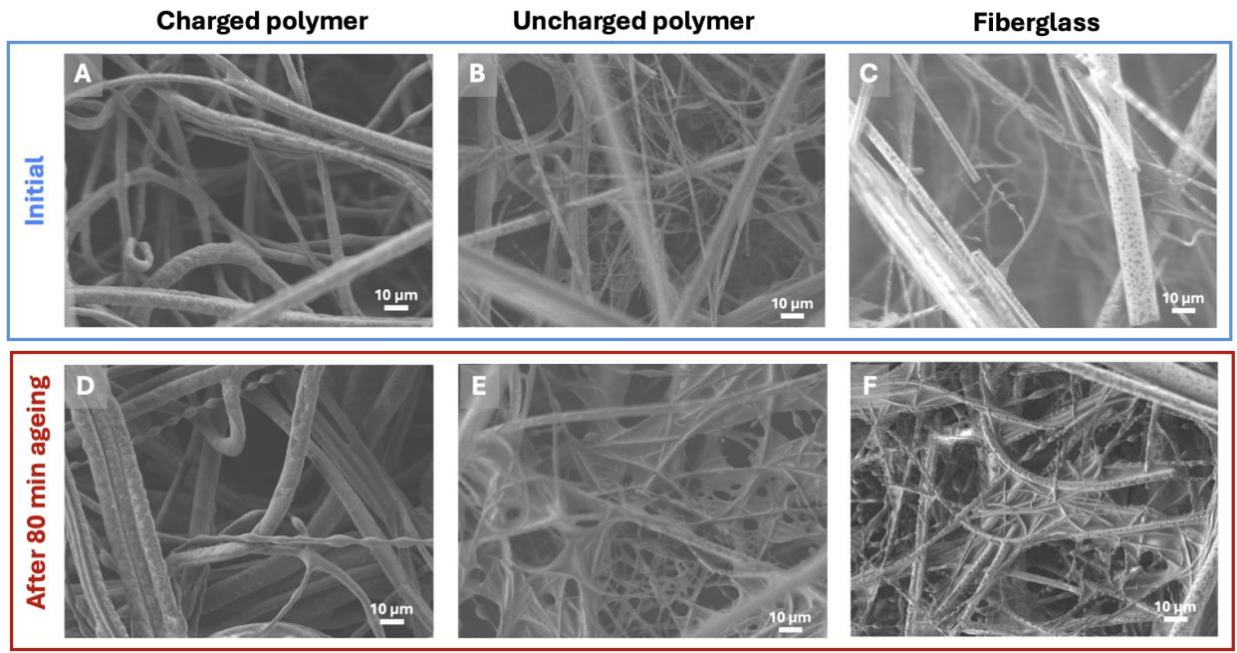

**Figure 3.** Representative Scanning Electron Microscopy (SEM) images of different media types rated MERV 13, captured after approximately 5 minutes (top row) and 80 minutes (bottom row) of smoke filtration. (A, D) Charged synthetic (polymer), Manufacturer C; (B, C) Uncharged synthetic (polymer), Manufacturer D; and (C, F) Fiberglass, Manufacturer D.

The extent and degree of beading and fiber coverage appear to be influenced by the charge; charged media exhibited much more extensive coverage than mechanical, uncharged media. There are three suggested mechanisms for the reduction of filtration efficiency for the charged air filters: 1) neutralization of charge by oppositely charged particles, 2) screening of the fiber charge by the captured particles, and 3) disruption of charge by dissolution or some chemical reaction taking place on the surface of the fiber.[24,25] Smoke particles, with their combination of solid (e.g., black carbon) and liquid/oil-like phases (e.g., SVOCs), present unique challenges. Some studies have shown that smoke particles carry a charge due to their carbonaceous component.

In summary, the effect of PN smoke on temporal filtration efficiency and the resulting PD on charged filter media differs significantly from that of inorganic particles loading. Smoke filtration's unique characteristics stem from specific interactions between PM smoke particles and media fibers, leading to distinct morphological features. Smoke deposition does not dramatically increase airflow resistance in any media types tested, a stark contrast to inorganic particle loading, indicating that an increase in PD may not be an effective diagnostic for recommending filter changes during wildfire smoke events. In addition, filters from different manufacturers with the same MERV rating can have varying smoke removal efficiencies, and in the case of charged media, higher MERV grades correlate with faster efficiency decay, necessitating frequent, unsustainable filter changes. A conclusion consistent with our previously published findings is that the standard recommendation to upgrade HVAC filters with higher MERV grades as a wildfire smoke mitigation strategy may not be effective. Simultaneously, mechanical filters' lower-than-expected efficiencies and high PD offer an energy-poor solution.

While small-scale lab testing of flat filter media offers valuable insights, testing larger flat-sheet and pleated filters at relevant flow rates is crucial, as the form factor and geometry can significantly affect smoke filtration performance. Even more pertinent, however, is testing under real-world conditions. Given the focus of many IAQ policies on HVAC filters, testing filters with smoke aerosols in real-world conditions should be a top priority to ensure public safety, enhance sustainability, and drive innovation in filtration.

**Supporting Information**: Additional experimental details including experimental procedures, photographs of the experimental setup, smoke characterization, filter media specs, and size-resolved filtration efficiency data. Information is mentioned in the text.

**Acknowledgment**

This work was performed in part at the Harvard University Center for Nanoscale Systems (CNS), a member of the National Nanotechnology Coordinated Infrastructure Network (NNCI), which is supported by the National Science Foundation under NSF award no. ECCS-2025158. This research was funded by the EPA SBIR Phase I Grant Number 68HERC24C0016. We thank Dr. Elijah Shirman for his help with the initial experimental setup.